\documentclass[journal ]{new-aiaa}
\usepackage[utf8]{inputenc}
\usepackage{textcomp}

\usepackage{graphicx}
\usepackage{amsmath}
\usepackage[version=4]{mhchem}
\usepackage{siunitx}
\usepackage{longtable,tabularx}
\title{Computationally examining the effect of plate thickness on hole emitter type electrospray thrusters}

\author{Sahil Maharaj \footnote{Postgraduate Researcher, Department of Mechanical, Aerospace and Civil Engineering, George Begg Building}, Mobin Yunus Malik \footnote{Postgraduate Researcher, Department of Mechanical, Aerospace and Civil Engineering, George Begg Building}, Olivier Allegre \footnote{Lecturer, Department of Mechanical, Aerospace and Civil Engineering, Pariser Building} and Katharine Lucy Smith \footnote{Reader(Associate Professor)
Department of Mechanical, Aerospace and Civil Engineering, George Begg Building}}
 \affil{University of Manchester, Oxford Rd, Manchester M13 9PL}

\begin{document}

\maketitle

\begin{abstract}
A new method for determining the onset voltage of electrospray thrusters is proposed, which specifically focuses on electrospray thrusters manufactured by laser drilling through flat plates. The novelty of this method is that it accounts for the effect of the thickness of the plate on the electrospray onset voltage requirements, while traditional methods do not. Key results from this study indicate  that for certain materials a change in thickness results
in a notable change in the onset voltage, which implies that the plate thickness needs to be considered  when planning the design of the thruster emitters.
This methodology allows for a robust method of observing the influence of key  parameters on the onset voltage. These developments can potentially facilitate  and improve the design of these thrusters, enabling an accurate understanding of the power requirements prior to  manufacture.
    
    
    
\end{abstract}

\section*{Nomenclature}

{\renewcommand\arraystretch{1.0}
\noindent\begin{longtable*}{@{}l @{\quad=\quad} l@{}}
$\beta$& Cone half angle\\
$\gamma$&Surface tension\\
$\epsilon_o$&Permittivity of vacuum\\
$\epsilon$&Relative permittivity\\
$\mu$&Viscosity\\
$\rho$&Density\\
$\sigma$&Electrical conductivity\\
$a$&Arbitrary Coefficient\\
$b$&Arbitrary Coefficient\\
$D$&Emitter diameter\\
$\mathbf{D}$& Electric displacement\\
$D_w$&Extractor diameter\\
$d$&Distance between emitter and extractor\\
$d_p$ & Distance to the extractor plane\\
$\mathbf{E}$& Electric field intensity\\
$E_o$ &Onset electric field strength\\
$E_r$ &Absolute Error\\
$f$&Equipotential surface factor\\
$g$&Standard acceleration due to gravity\\
$h$&Capillary height\\
$I$&Current\\
$\frac{\partial I}{\partial v}$& Volumetric current\\
$\mathbf{J}$& Current density\\
$K$&Conductivity\\
$n$&Normal pointing away from domain\\
$n_e$&Number of elements\\
$Q_{min}$&Minimum flow rate at which ES is possible\\
$r_m$&Radius of meniscus\\
$r_t$&Radius at apex of Taylor cone\\
$t$&Thickness of plate\\
$V$&Scalar electric potential\\
$V_o$ &Onset voltage\\
$V_of$ &Onset voltage (Hole Emitter)\\
$V_b$&Breakdown voltage\\
$v$&Volume\\
$v_s$&Separation constant\\
$w$&Capillary width\\
\end{longtable*}}

\section{Introduction}
\lettrine{A}{mongst} the various options for on-board propulsion systems available to satellite designers, few are small enough to fit onto a nano-satellite. With the rise in popularity of nano and pico-satellites, driven by the propagation of the CubeSat standard, this has become a niche that needs to be filled \citep{lemmerPropulsionCubeSats2017}. CubeSats present a way for smaller organisations to design and launch a satellite in a significantly easier and more cost-effective manner than previously would have been possible. As such, an easily available source of onboard propulsion would improve the effectiveness of any such satellites produced, allowing them to be used for missions with longer lifetimes, or at lower altitudes. This would increase the scope of missions that these satellites can undertake, which would ideally lead to a higher uptake in their use.

While various technologies are being examined to fill this niche, this paper focuses solely on electrospray (ES) propulsion. This is a form of electrostatic propulsion that can produce thrusts in the  \SI{}{\micro \newton} range, and specific impulses of over \SI{1 000}{\second} \citep{dandavinoMicrofabricatedElectrosprayEmitter2014}, using emitters with volumes and masses low enough to be compatible with CubeSat platform constraints. Thrust is achieved through the use of electric fields acting on a conductive propellant, which in systems requiring high specific impulse in vacuum conditions is typically an ionic liquid \citep{millerCapillaryIonicLiquid2021}. The propellant is fed into an emitter, which has a potential difference applied between it and an extractor electrode. This potential difference generates the electric field, which causes the propellant to be attracted toward the electrode. At first, the propellant will deform into a cone shape, known as a Taylor cone \citep{delamoraCurrentEmittedHighly1994a}, before beginning to emit charged particles as the electric field strength increases. The emission of these accelerated charged particles results in a through force acting on the spacecraft, propelling the system in the opposite direction of the emission. The magnitude of the potential difference at which this spray emission begins to be observed is referred to as the onset voltage ($V_o$), which is one of the more important parameters when designing an ES system. Minimising the $V_o$ reduces the proportion of output of the power supply dedicated to obtaining the initial electrospray, which implies that more power will be available for other uses. Typically, in an electrospray system another electrode, known as the accelerator electrode, will be connected to a separate accelerator circuit. This is due to the fact that in order for the spray to remain stable there is a limited range of potential differences at which the extractor electrode can be maintained. However, a greater potential difference would serve to increase the velocity at which the particles travel \citep{dandavinoMicrofabricatedElectrosprayEmitter2014}. This electrode is thus used to further accelerate the particles emitted, to improve the performance of the system, as higher velocity particles result in a higher thrust and specific impulse. If less power output is being allocated to the extractor circuit, this means a greater proportion of the onboard power is available for the accelerator circuit, or, alternately, to other systems on board.

Traditionally, the emitter in an electrospray system consists of a needle or capillary shape. This is due to the fact that these shapes are able to focus the electric field at the point where the Taylor cone forms, thus reducing the onset voltage. In these cases, the method for empirically determining the value of $V_o$ is given by the equation \ref{eq:onset_voltage_long} \citep{krpounMethodDetermineOnset2008}. Here, $\gamma$ is the surface tension of the propellant, $D$ is the diameter of the emitter, $\epsilon_o$ is the permittivity of vacuum and $d$ is the distance between the extractor and the emitter.

\begin{equation}
    \label{eq:onset_voltage_long}
     V_o=\sqrt{\frac{\gamma  D}{2 \epsilon_o}}\frac{ln(\frac{D+4d+4\sqrt{d(d+\frac{1}{2} D)}}{D})}{\sqrt{1+\frac{D}{2d}}}
\end{equation}

However, this standard capillary emitter is not the only possible layout for an electrospray thruster. Another possible layout involves externally wetted emitters, where the propellant covers cone-shaped structures, with the Taylor cone forming at the apex. This geometry can also be combined with the capillary emitter, by having the propellant be transported to the emitter by capillaries \citep{siegel_silicon_2020}. Additionally, a layout that is less developed, but which has been the significant focus in more recent research are porous emitters \citep{natisinFabricationCharacterizationFully2020}. These emitters involve the propellant being fed through a porous substrate which is shaped to provide a focused electric field to support electrospray formation. This process  \citep{leggeElectrosprayPropulsionBased2011} typically results in lower flow rate electrosprays, which are desirable to obtain ionic emission, the advantages of which are discussed later in this paper.

One other possibility is a variation of the capillary emitter which involves constructing an emitter by drilling a hole in a flat plate of material, this layout is the focus of this paper. This hole emitter supplies fluid to the Taylor cone, similar to a capillary emitter, with the propellant being fed through the hole towards the extractor. There have been some investigations into manufacturing and testing similar geometries \citep{paine_micro-fabricated_2001}, with manufacturing being time-consuming and requiring multiple pieces of equipment. However, preliminary work indicates that with recent developments in laser manufacturing technology, these emitters may now be quicker and easier to manufacture than other types of emitters \citep{maharaj_photon_2021}. While work has been done on similar systems using traditional CNC manufacturing methods \citep{kimber_dielectric_2018}, laser manufacturing the system will allow for smaller features, which both improve performance and allows for the entire system to be more compact. Laser manufacturing is a field that has the potential to significantly impact the process of electrospray emitter manufacture, with the benefits of this area already being subject to ongoing research.  As an example of this, research 
 by \citet{terekhovUltrafastLaserFabrication2021} has demonstrated that laser ablation can be used to generate emitters out of a fused silica chip. This is similar to the research described in this paper, however, this research focuses on drilling through plates of low permittivity dielectric materials. This has the potential to decrease the costs of the system, as these materials have greater availability, particularly in light of contemporary shortages of silicon \citep{yangInvestigationNa2CO3CaO2021}.

As with capillary emitters, the relationship between the various geometric and material parameters of these emitters and the onset voltage can be estimated based on empirical analysis. Experimental investigations into the performance of these emitters have found that they can achieve ES emission at voltages that are comparable to, or lower than, the onset voltages for capillary emitters \citep{lozano_flatplate_2004}, with the analytically derived equation for the onset voltage in a hole emitter (denoted by $V_{of}$) given by equation \ref{eq:onset_voltage_hole}. Here, $r_m$ is the radius of the meniscus, $f$ is the factor that determines the equipotential surface of the cone and $v_s$ is the separation constant.

\begin{equation}
    \label{eq:onset_voltage_hole}
     V_{of}=2\sqrt \frac{\gamma r_m}{\epsilon_o} \frac{(d/r_m)^{v_s}+f^{-2v_s-1}-1}{v_s+(v_s+1)f^{-2v_s-1}}
\end{equation}

It should be noted that this expression does not take into account the thickness of the plate that was drilled into to make the emitter. Previous work \citep{maharaj_propulsion_2021} has established that the thickness of the plate has an effect on the magnitude of the peak electric field, with thicker plates resulting in a higher electric field strength for some materials. As the field strength determines the onset voltage, this implies that thicker plates would result in lower onset voltages, a factor that is not reflected in the analytically devised equation. 

This is important as thicker plates have a variety of drawbacks. Thicker plates require more time to machine, especially if an emitter with multiple holes is required. Additionally, when laser drilling an emitter, a thicker plate will result in the taper angle of the hole having a more pronounced effect, resulting in greater deformation of the hole geometry. Finally, a thicker plate will have a higher mass than a thinner plate, with even a small increase being significant on a small satellite. Considering these points, it can be seen that if the basic analytical expression, that does not account for plate thickness, is used, then an infinitely thin plate is ideal. However, if the thickness does have a beneficial effect on the onset voltage, then this will present a trade-off, wherein the benefits of the lower onset voltage will need to be compared to the detriments that come with an increase in thickness. The only exception to this comes when considering the operating regime of the emitter. Electrospray emitters can operate in the purely ionic regime, where ions are emitted, the droplet regime, where larger droplets of liquid are emitted, or a combination of the two. The aspect ratio of the emitter has an effect on the regime \citep{siegel_silicon_2020}, with a hole with a radius of \SI{45}{\micro \meter} requiring a thickness of over \SI{6.5}{\milli \meter} to approach the purely ionic regime. However, this will not be considered in this paper, as this presents a more subtle influence on thruster performance that would require a more detailed mission analysis beyond the scope of this paper. While both regimes have different operating characteristics, thrusters operating in the droplet regime, or emitting a combination of both, still have all the advantages identified earlier. For a recent example, NASA’s Disturbance Reduction System (DRS) mission (which was integrated into ESA's LISA pathfinder spacecraft) utilised an emitter in the droplet mode due to the stable and precise nature of this mode \citep{demmons_electrospray_nodate}, indicating that there is still a need for thrusters operating in this regime.
Having established this, the aim of this paper is to determine the effect of the thickness of the plate on the onset voltage for hole emitters made of different materials. This can then be used in the future to estimate the minimum necessary plate thickness for a given power supply, in order to minimise production time, cost, and thruster mass.
 
The results presented were obtained using computational modelling, as this allowed a greater degree of freedom in adjusting the various parameters, as well as quickly and easily measuring the electric field strength at various points. More specifically, the geometry of an emitter was modelled in COMSOL multiphysics \citep{comsol_ab_comsol_nodate}, a multiphysics simulation platform and finite element analysis solver. Different geometries were modelled, and the potential difference was varied until an electric field strength suitable for ES emission was reached. The magnitude of these suitable electric field strength values was determined by two methods. Firstly, by taking equations from literature that were analytically derived, and secondly, by modelling a capillary emitter with the potential difference set to the known onset voltage and measuring the electric field strength. In lieu of performing new experimental work, the results were verified by modelling a system based on previous experimental work done by another research group and comparing it to the results described in literature. This research is similar to that undertaken by \citealt{jonesNumericalInvestigationEffects2020}, who examined a similar method of using numerical modelling to investigate the effect of the thickness of a plate on the onset voltage. However, unlike that study, this investigation shows the potential to easily adjust other parameters in order to determine their effects on the onset voltage, such as the taper and wetting. This method thus has the potential to be more versatile when designing emitters with suitable geometries.  

It should be noted that this paper is describing the onset voltage of a single emitter, with the effect of multiple emitters in a capillary being the subject of future work. Research \citep{krpounMethodDetermineOnset2008} has shown that the spacing of emitters in an array can have an effect on the onset voltage, due to the interaction between the electric fields, so the spacing between emitters would need to be carefully considered. However, that would need to be considered when designing a complete thruster system, and as this paper is focused on the manufacturing of an emitter it will be considered beyond the scope of this project. 

\section{Computational Simulation}

\begin{figure}[h]
    \centering
    \includegraphics[width=0.9\textwidth]{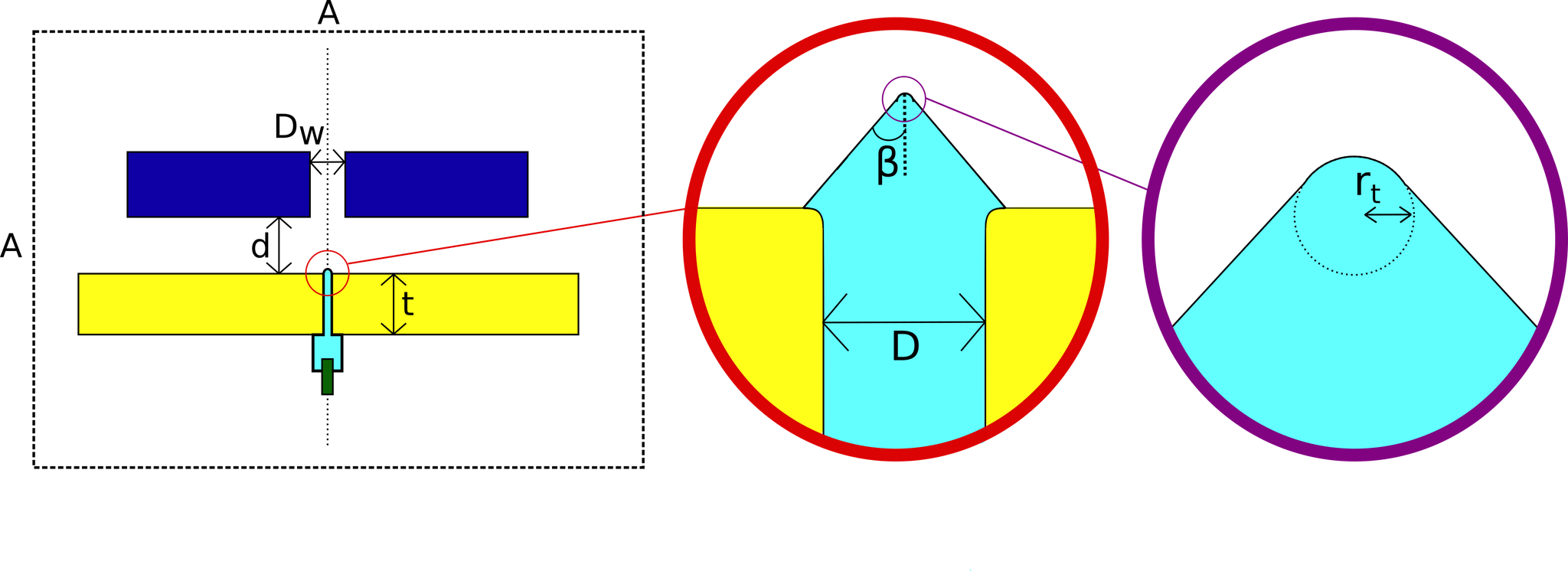}
    \caption{Representation of flat plate geometry as computationally modelled. Adapted with permission from \citet{maharaj_photon_2021}}
    \label{fig:COMSOL_geometry}
\end{figure}

Figure \ref{fig:COMSOL_geometry} depicts a cross-section of the geometry that was modelled. Here, the yellow shape represents a flat plate, with a hole in center filled with propellant. The propellant is capped with a Taylor cone, with the red circle depicting a zoomed-in view of the cone itself. The tip of the cone is a spherical cap tangential to the cone, with the purple circle depicting a further zoomed-in version of the tip of the cone. The green and dark blue shapes represent the electrodes, which will be used to generate the electric field. Finally, the dotted square around the geometry represents the modelled domain. It should be noted that this model does include some simplifications compared to an actual experimental rig. In a typical experimental rig, there will be a propellant reservoir that feeds the liquid to the emitter, with the emitter electrode being introduced between the reservoir and the emitter. However, initial investigations into modeling this system found that simplifying the system by removing the reservoir and simplifying the design of the electrode did not affect results, but allowed for the meshing of the system to take significantly less time.

\begin{table}[h]
    \centering
        \caption{Summary of parameters used when modelling the hole and capillary emitters}
    \begin{tabular}{c c c c}
    \hline
         Parameter Name & Symbol & Fixed Value & Model Used  \\
        \hline \hline
        Diameter of Extractor&$D_w$&\SI{0.7}{\milli \meter}&Flat Plate\\ \hline
         Radius at Tip&$r_t$&\SI{0.2}{\micro \meter}&Both\\ \hline
         Taylor Cone Half Angle&$\beta$&\SI{49.3}{\degree}&Both\\ \hline
        Hole Diameter&$D$&\SI{45}{\micro \meter}&Both\\ \hline
        Plate Thickness&$t$&Varied&Flat Plate\\ \hline
        Distance to Extractor&$d$&Varied&Both\\ \hline
        Capillary Width&$w$&\SI{0.1}{\milli \meter}&Capillary\\ \hline
        Capillary Height&$h$&\SI{7}{\milli \meter}&Capillary\\ \hline
         Length of Computational Domain&$A$&\SI{50}{\centi \meter}&Capillary\\ \hline
         
    \end{tabular}

    \label{tab:parametersummary}
\end{table}

Figure \ref{fig:COMSOL_geometry} also displays the important geometric parameters that need to either be adjusted or set to a fixed value. All of these values are described in table \ref{tab:materials}, along with some parameters that are used in the capillary model that will be described later. For all experiments, the modelled domain had a length and breadth of \SI{50}{\centi \meter}, with the described geometry located in the center. $D_w$, the diameter of the extractor, was set to \SI{0.7}{\milli \meter} for all simulations, while $r_t$, the radius of the tip, was always set to \SI{0.2}{\micro \meter}. This value for the radius at the tip was obtained using a formula that describes the relationship between the flow rate and radius, which is given in equation \ref{eq:radius_at_tip}, taken from \citet{gamero-castano_direct_2000}. Here the flow rate $Q$ is taken as the minimum flow rate. 
\begin{equation}
     r_t={\frac{\epsilon_o Q}{K}}^{1/3}
     \label{eq:radius_at_tip}
\end{equation}


The Taylor cone itself was modelled as an ideal Taylor cone, which meant that the half angle $\beta$ was kept constant at \SI{49.3}{\degree}. The properties of the hole itself were modelled on a hole that was produced experimentally \citep{maharaj_photon_2021}. This hole had a diameter $D$ of \SI{45}{\micro \meter}, and while the experimentally manufactured holes were tapered, this was ignored to make comparison easier. Additionally, it should be noted that the cone was modelled as having some wetting across the surface, such that the radius of the cone base was \SI{17}{\micro \meter} wider than the radius of the emitter. All other geometric parameters described by Fig. \ref{fig:COMSOL_geometry} (namely the thickness $t$ and the distance $d$) are varied across the tests run. It should be noted that the propellant was modelled as a triethylene glycol (TEG) and sodium iodide (NaI) solution, due to the fact that previous testing in atmosphere was done using such a solution as a propellant. The material properties of this propellant, as well as every other component of the model, can be seen in table \ref{tab:materials}.

\begin{table}[h]
    \centering
        \caption{Summary of Material Properties}
    \begin{tabular}{c c c c}
    \hline
         Component & Material & Conductivity (\SI{}{\siemens \per \meter}) & Relative permittivity\footnote{Relative permittivity is unitless}  \\
        \hline \hline
        Propellant & TEG/NaI mixture&0.04&23.69\\ \hline
        Electrodes & Aluminium&\SI{3.77e7}{} &1\\ \hline
        Emitter Plate & PTFE&\SI{1e-24}{}&2.1\\ \hline
        Surrounding medium & Air&\SI{6e-15}{}&1\\ \hline

    \end{tabular}

    \label{tab:materials}
\end{table}

 In order to simulate the electric current across the geometry, a stationary, asymmetric model was constructed based on two constitutive equations. These can be seen in equations \ref{eq:consititutive_1} and \ref{eq:consititutive_2}  
 
     \begin{equation}
  \mathbf{J}=\sigma \mathbf{E}
     \label{eq:consititutive_1}
 \end{equation}
 
    \begin{equation}
  \mathbf{D}=\epsilon_0 \epsilon \mathbf{E}
     \label{eq:consititutive_2}
 \end{equation}
 
 In these equations, $\mathbf{D}$ is the electric displacement, $\mathbf{E}$ is the electric field intensity, $\mathbf{J}$ is the current density and $\sigma$ is the electrical conductivity.
 Within the model $\epsilon_o$ was defined as having a value of \SI{8.85e-12}{\farad \per \meter}, and the values of $\sigma$ and $\epsilon$ are determined by the material properties presented in table \ref{tab:materials}. The primary conservation equations can be seen in equations \ref{eq:conservation_1} and  \ref{eq:conservation_3}, with equation \ref{eq:conservation_1} being derived from Ohm's Law, and equation \ref{eq:conservation_3} defining the electric field in terms of the electric potential 
 
 \begin{equation}
   \mathbf{\nabla \cdot J}=\frac{\partial I}{\partial v}
     \label{eq:conservation_1}
 \end{equation}
 
 
     \begin{equation}
  \mathbf{E}=-\nabla V 
     \label{eq:conservation_3}
 \end{equation}
 
 In these equation V is the scalar electric potential and $\frac{\partial I}{\partial v}$ represents a volumetric source of current (measured in \SI{}{\ampere \per \meter \cubed}). 
 
 The boundary conditions between interior boundaries were defined by equation \ref{eq:boundary_conditions}, where $\mathbf{n_1}$ is the normal pointing away from the domain of a given material, and $\mathbf{J_1}$ and $\mathbf{J_1}$ are the total current density of both domains. This equation is used to ensure current continuity.
 
 
  \begin{equation}
  \mathbf{n_1 \cdot (J_1-J_2)}=0
     \label{eq:boundary_conditions}
 \end{equation}
 
Only half of the geometry was modelled, with the dotted line in figure \ref{fig:COMSOL_geometry} being treated as a line of axial symmetry. All other exterior boundaries were treated as being electrically insulated, according to equation \ref{eq:insulation} \citep{acdc_users_manual}.

\begin{equation}
   \mathbf{n \cdot J}=0
     \label{eq:insulation}
 \end{equation}

While this model was the primary one used, a second model needed to be created for a capillary emitter, which was used to validate the electric field at which ES emission would be observed. This model was similar to the previously described model, except with some notable differences. In the flat plate model, the plate has a given thickness $t$, while the width of the plate is wide enough compared to the hole that it can be treated as having an infinite width. In comparison, the capillary emitter has a set width ($w$), which in this case was set to \SI{0.1}{\milli \meter}, while the height of the needle ($h$) (equivalent to the thickness $t$ of the plate) was set to \SI{7}{\milli \meter}. These values are also listed in table \ref{tab:parametersummary}. 
This capillary was treated as being made of stainless steel, with a conductivity and relative permittivity of \SI{4.03e6}{\siemens \per \meter} and \SI{1}{} respectively. Stainless steel was chosen due to its status as a pure conductor, which would give field conditions similar to most other metals, with metal capillaries being used in initial experimental testing. The propellant was modelled as the same TEG/NaI mixture described in table \ref{tab:materials}. The capillary model was used to determine the accuracy of the onset electric field value obtained from equation \ref{eq:onset_field}, by modelling the system at known onset voltages and then measuring the generated electric field strength.
 
 \section{Onset Voltage Model}
 
 As the results of these simulations give the peak electric field strength, it was required to establish how this related to the onset voltage. The electric field strength at which ES emission would begin to be observed ($E_o$) was described by \citet{mair_emission_1980}, using the formula described by equation \ref{eq:onset_field}. 
 
 \begin{equation}
     E_o=(\frac{4 \gamma}{\epsilon_o r_t})^{1/2}
     \label{eq:onset_field}
 \end{equation}
 
 In this case, this equation indicates that ES emission will begin to be observed when the peak electric field strength is \SI{3.2e8}{\volt \per \meter}. This means that the onset voltage can be calculated by modelling the system with a range of potential differences, and noting the point at which the peak electric field strength reaches the value obtained from equation \ref{eq:onset_field}. The first set of tests were done in order to confirm this value, while also verifying the computational model. This involved modelling the capillary setup and setting the potential difference to the known onset voltages. These onset voltage values were obtained from equation \ref{eq:onset_voltage_long}, which has a history of use in experimental work \citep{lajhar_electrospray_2018}.
 
\begin{figure}[h!]
    \centering
    \includegraphics[width=0.9\textwidth,trim={0.1cm 0.1cm 0.1cm 0.1cm},clip]{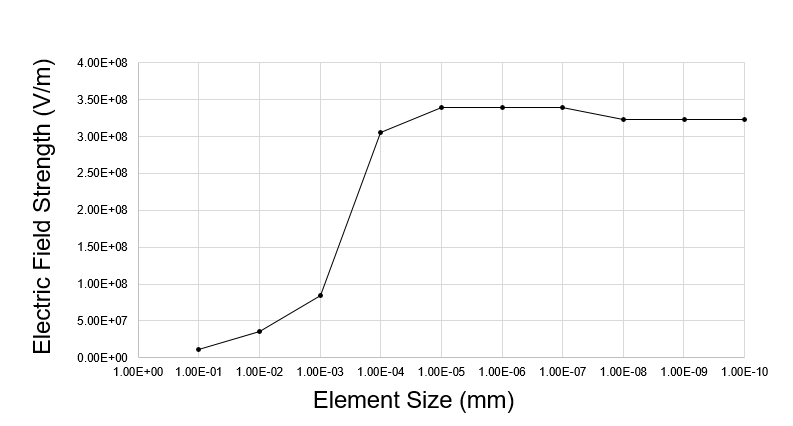}
    \caption{Electric field strength data from mesh refinement study}
    \label{fig:mesh_refinement}
\end{figure}

 However, before this testing could be undertaken, a mesh refinement study needed to be conducted. This involved modelling the capillary emitter with the distance between the emitter and extractor ($d$) set to \SI{1.056}{\milli \meter}, and the voltage set to  \SI{1779.7}{\volt}, which was obtained from equation \ref{eq:onset_voltage_long}. 
 This setup was modelled, and the electric field at the tip of the cone was measured multiple times with a progressively finer unstructured free triangular mesh each time, measured by the minimum element size of the mesh. The results of this can be seen in Fig. \ref{fig:mesh_refinement}. For the purposes of readability, the x-axis was inverted.

\begin{figure}[h!]
    \centering
    \includegraphics[width=0.9\textwidth,trim={0.1cm 0.1cm 0.1cm 0.1cm},clip]{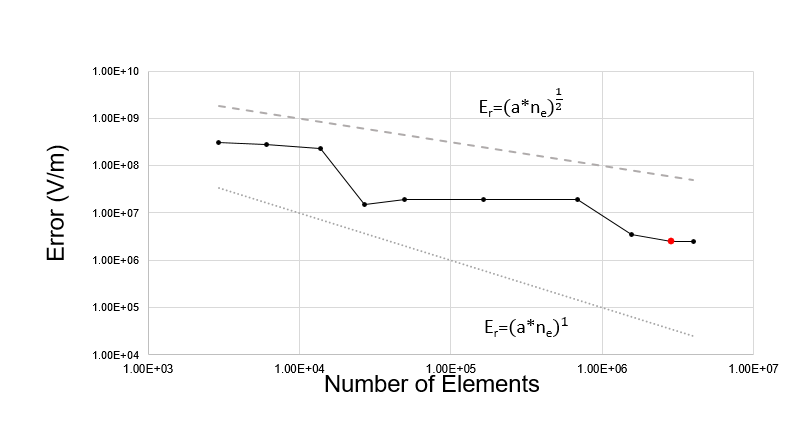}
    \caption{Absolute error estimation data from mesh refinement study, with the dotted and dashed lines provide for the sake of comparison. The red dot represents the point at which the results converge}
    \label{fig:mesh_ErrorEstimate}
\end{figure}

 Here it can be seen that as the minimum element size decreases, the measured electric field strength increases until it reaches a magnitude of \SI{3.37e8}{\volt \per \meter}, at which point it maintains a similar value until the element size is decreased to \SI{1e-8}{\milli \meter}. At this point the value of the electric field decreases to \SI{3.24e8}{\volt \per \meter}, with it maintaining a similar value as the mesh is further refined. It's notable that this graph does not demonstrate a linear convergence, which might be an indication that there are competing effects within the model being simulated. This can be further examined by examining error estimations. As previously explained, this is a computational model of a real system wherein the electric field strength can be computed using equation \ref{eq:onset_field}. As such, the error in the model can be estimated by comparing the output of the model, to the result obtained from the equation\footnote{This does assume that both Equation \ref{eq:onset_field} and \ref{eq:onset_voltage_long} are accurate. However, evidence that these are valid assumptions is provided later in this section.}. Figure \ref{fig:mesh_ErrorEstimate} shows the absolute difference between the two values of electric field strength ($E_r$), compared to the number of elements in the mesh ($n_e$). The dotted and dashed lines in the figure are provided for comparison's sake. The dotted line is to the first order, and the dashed line is to the order $\frac{1}{2}$. Note that the lines are multiplied by arbitrary coefficients a and b (which are \SI{1e11}{} and \SI{1e10}{} respectively). This was done solely for the purpose of readability, as it helped position the lines close to the error line they are being compared to. The slope of the error line more closely matches the dashed line, which is to the order $\frac{1}{2}$.  It can thus be said that the mesh refinement has less than first-order convergence. 

\begin{figure}[h!]
    \centering
    \includegraphics[width=0.9\textwidth,trim={0.1cm 0.1cm 0.1cm 0.1cm},clip]{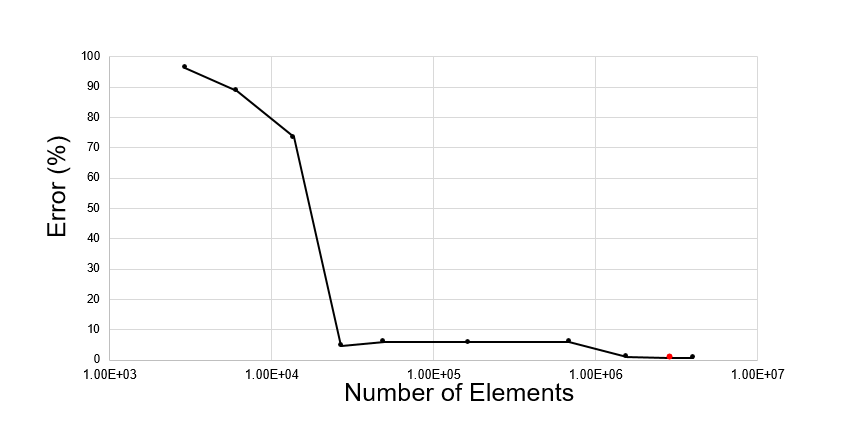}
    \caption{Error estimation data from mesh refinement study expressed as a percentage of the expected value}
    \label{fig:mesh_Error_Percentage}
\end{figure}

 Figure \ref{fig:mesh_Error_Percentage} shows this as a percentage of the expected value. Here we can see that the result converges with the analytical solution at the point that is marked with a red dot on both \ref{fig:mesh_ErrorEstimate} and \ref{fig:mesh_Error_Percentage}. This corresponds with a minimum element size of \SI{1e-9}{\milli \meter}, and an error value of \SI{1.5e7}{\volt \per \meter}. This value was only a 0.8\% change from the analytically derived value. 
 
\begin{figure}[h!]
    \centering
    \includegraphics[width=0.3\textwidth]{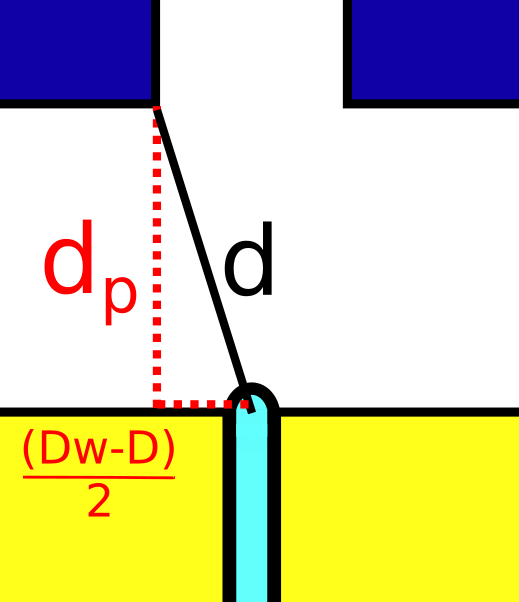}
    \caption{Method for calculating distance to extractor}
    \label{fig:distance_pythag}
\end{figure}

  With the information obtained from the mesh refinement study, the mesh for all future tests was generated with a minimum element size of \SI{1e-9}{\meter}. Once the mesh refinement study was done, it was possible to return to examining the onset electric field strength. The distance between the emitter and extractor was varied, in order to obtain the onset field strength under a range of conditions. The distances were set to \SI{1.056}{\milli \meter}, \SI{2.112}{\milli \meter}, \SI{3.076}{\milli \meter} and \SI{4.057}{\milli \meter}, with these values selected by adjusting the distance to the extractor plane, and determining the true distance using the Pythagorean theorem. A diagram clarifying this process can be seen in Fig. \ref{fig:distance_pythag}. Here, $d_p$ is the distance to the extractor plane.
 
 For each value of $d$, the onset voltage was set to values obtained from equation \ref{eq:onset_voltage_long}. The potential difference here (and indeed in all future tests) was generated by applying a fixed potential across the boundaries of the electrode connected to the emitter and grounding the boundaries of the extractor electrode. The potential difference values for this first test, as well as the corresponding peak field strength, can be seen in table \ref{tab:test1}. The electric field strength results average out to \SI{3.19e8}{\volt \per \meter}, with a standard deviation of \SI{0.028e8}{\volt \per \meter}. The average figure is only a 0.4\% decrease from the value obtained in equation \ref{eq:onset_field}, which is a negligible change. As these two methods complement each other, it indicates that the obtained electric field value is an appropriate one, and as such, this value will be used in all future tests.
 
  \begin{table}[h]
     \centering
          \caption{Results of capillary test}
     \begin{tabular}{c c c}
     \hline
        Distance to extractor (\SI{}{\milli \meter})  & Potential Difference (\SI{}{\volt})& Electric Field Strength (\SI{}{\volt \per \meter}) \\
        \hline \hline
         1.056 & 1779.7 & \SI{3.23e8}{}\\ \hline
         2.112& 2005.4 & \SI{3.20e8}{}\\ \hline
         3.076 & 2104.3 & \SI{3.16e8}{}\\ \hline
         4.057 & 2236.7 & \SI{3.16e8}{}\\ \hline
     \end{tabular}
     \label{tab:test1}
 \end{table}
 
 In order to further validate the computational model, a flat plate model with a known onset voltage was modelled. This was based on an experimental test done by \citet{lozano_flatplate_2004}, using a \SI{7}{\milli \meter} thick sheet of polytetrafluoroethylene (PTFE). Here, the extractor was set \SI{3.5}{\milli \meter} from the hole emitter, which had a radius of \SI{166.5}{\micro \meter}. The potential difference between the two was set to \SI{2.6}{\kilo \volt}, which was the experimentally observed onset voltage, and the simulated electric field strength was \SI{3.07e8}{\volt \per \meter}. This is 96.3\% of the mean simulated value and almost a perfect match to the value obtained from equation \ref{eq:onset_field}. This further indicates that the simulation is able to correctly model the electric field strength at the onset of ES emission. With this in mind, all further tests simulate the model across a range of potential differences, until the generated electric field strength approaches \SI{3.2e8}{\volt \per \meter} (which will be referred to as the target voltage).
 
 \section{Effect of plate thickness}
 
 The previously described method was used to investigate the effect of plate thickness on onset voltage for emitters constructed out of two different materials. The first of these was polytetrafluoroethylene (PTFE), as this was the material that was described as having the lowest onset voltages by \citet{lozano_flatplate_2004}. However, in previous investigations, this material did not exhibit a notable change in electric field strength when the material thickness was varied \citep{maharaj_propulsion_2021}. As such, the next material chosen was polyether ether ketone (PEEK). This was selected as it generated electric field strength values comparable to PTFE at higher thicknesses, but unlike PTFE experienced a significant change in peak electric field values as the thickness changed. The relevant material properties that affected the strength of the electric field are the conductivity and the permittivity of the plate. The relative permittivity of PTFE was set as \SI{2.1}{}, and the conductivity was set as \SI{1e-23}{\siemens \per \meter}, while the permittivity and conductivity of PEEK was set as \SI{3.3}{} and \SI{2.04e-15}{\siemens \per \meter} respectively.

 For both materials, the geometry was modelled as described in Fig. 1, with the distance to the extractor set to \SI{1.056}{\milli \meter}. The thickness of the plate was varied from \SI{1}{\milli \meter} to \SI{7}{\milli \meter}, and the potential difference was iteratively
adjusted until appropriate values were reached. In order to simplify this stage of testing, the potential difference was
varied in steps of \SI{10}{\volt}. As the expected results were in the kiloVolt magnitude, this can introduce a margin of error in the range of 1\%. This amount is negligible, as it can be compared to noise fluctuations on most power sources. However, uncertainty induced by this step size will still be taken into account when examining the results.
 
For the PTFE sheet, the voltages were varied between \SI{800}{\volt} and \SI{900}{\volt} for all thicknesses. The results of this test can be seen in table \ref{tab:PTFE_results}. This includes the thickness of the plate, the measured voltage, the electric field strength that was generated, and finally the absolute difference between the generated voltage and the target voltage. The absolute difference is intended to give an idea of the margin of error, with lower values indicating a closer match to the true onset voltage.

\begin{table}[h]
    \centering
        \caption{Results of PTFE test}
    \begin{tabular}{c c c c}
    \hline
        Plate Thickness (\SI{}{\milli \meter}) & Electric Field Strength (\SI{}{\volt \per \meter})& Absolute Difference (\SI{}{\volt \per \meter}) & Measured Onset Voltage (\SI{}{\volt})\\
\hline \hline
        1 & \SI{3.201e8}{}&\SI{0.100e6}{}&830 \\ \hline
        2 & \SI{3.205e8}{}&\SI{0.500e6}{}&850 \\ \hline
        3 & \SI{3.206e8}{}&\SI{0.600e6}{}&830 \\ \hline
        4 & \SI{3.185e8}{}&\SI{1.500e6}{}&830 \\ \hline
        5 & \SI{3.203e8}{}&\SI{0.300e6}{}&850 \\ \hline
        6 & \SI{3.202e8}{}&\SI{0.200e6}{}&850 \\ \hline
        7 & \SI{3.188e8}{}&\SI{1.200e6}{}&840 \\ \hline
    \end{tabular}

    \label{tab:PTFE_results}
\end{table}
 
 In contrast to the small spread of potential difference values for PTFE, the PEEK test was done with a range of values from \SI{1000}{\volt} to \SI{1250}{\volt}. Other than the material change, all other parameters were kept the same as the PTFE test, with the results presented in table \ref{tab:PEEK_results}.
 
 \begin{table}[h]
    \centering
        \caption{Results of PEEK test}
        \label{tab:PEEK_results}
    \begin{tabular}{c c c c}
        \hline
        Plate Thickness (\SI{}{\milli \meter}) & Electric Field Strength (\SI{}{\volt \per \meter})& Absolute Difference (\SI{}{\volt \per \meter}) & Measured Onset Voltage (\SI{}{\volt})\\
         \hline\hline
        1 & \SI{3.215e8}{}&\SI{1.500e6}{}&1220 \\
        \hline
        2 & \SI{3.192e8}{}&\SI{0.800e6}{}&1140 \\
        \hline
        3 & \SI{3.204e8}{}&\SI{0.400e6}{}&1100 \\
        \hline
        4 & \SI{3.190e8}{}&\SI{1.000e6}{}&1090 \\
        \hline
        5 & \SI{3.190e8}{}&\SI{1.000e6}{}&1080 \\
        \hline
        6 & \SI{3.178e8}{}&\SI{2.230e6}{}&1070 \\
        \hline
        7 & \SI{3.190e8}{}&\SI{1.050e6}{}&1060 \\
        \hline
    \end{tabular}

\end{table}

\begin{figure}[h]
    \centering
    \includegraphics[width=0.7\textwidth,trim={0.1cm 0.1cm 0.1cm 0.1cm},clip]{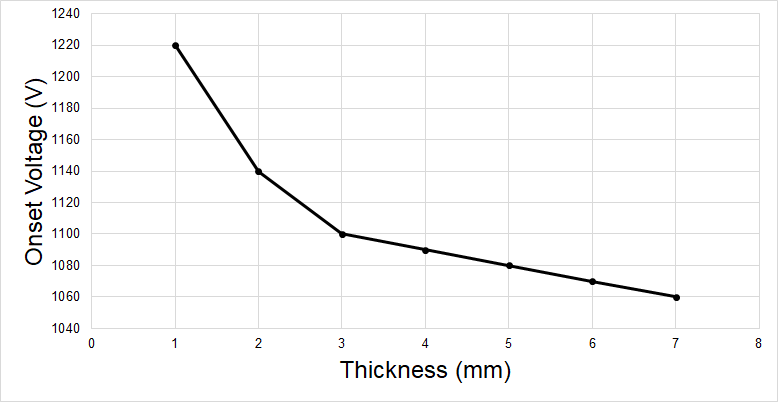}
    \caption{Change in onset voltage across PEEK emitters of different thicknesses}
    \label{fig:PEEK_onset_voltage_1}
\end{figure}

Referring to these results, it can be seen that unlike the PTFE test, where the onset voltage seems to fluctuate between \SI{850}{\volt} and \SI{830}{\volt}, the PEEK emitter onset voltage decreases as the thickness is increased. Figure \ref{fig:PEEK_onset_voltage_1} shows this trend, wherein it can be seen the differences are more pronounced at first before the results begin to level out. However, despite this trend, there is still a decrease of \SI{160}{\volt} in the onset voltage as the plate thickness increases from \SI{1}{\milli \meter} to \SI{7}{\milli \meter}. This is a meaningful change, as it corresponds to a 13\% drop in the original onset voltage.

\section{Sources of error and uncertainty}

In both cases described in the previous section, the margin of error introduced by the voltage rounding should not have enough of an effect to cast doubt on these results. For PTFE, the highest absolute difference value was \SI{1.5e6}{\volt \per \meter}, which was less than 0.5\% of the target electric field strength of \SI{3.2e8}{\volt \per \meter}. Similarly, the maximum difference for PEEK is \SI{2.23e6}{\volt \per \meter}, which corresponds to 0.7\% of the desired value. Both of these figures are small enough that they can be considered negligible.

It should be noted that for both cases, the onset electric field strength was set to the value obtained from equation \ref{eq:onset_field} (\SI{3.2e8}{\volt \per \meter}), rather than the average figure obtained from the simulation (\SI{3.19e8}{\volt \per \meter}), rather than the figure obtained. This was done due to the fact that this is the larger figure, and as such results in an extra factor of safety. As it will be possible to achieve electrospray if the maximum generated electric field strength exceeds the onset field strength, but not if it doesn't, using the higher values decreases the chances of falsely concluding that a setup is viable. However, this is a very minor deviation and is unlikely to have any effect on the results.

The degree to which this impacts the results can easily be measured. Replicating the test on the first row of table \ref{tab:PEEK_results}, but aiming for an electric field strength of \SI{3.19e8}{\volt \per \meter}, gives an electric field of \SI{3.15}{\volt \per \meter} at \SI{1210}{\volt}. This is a 0.8\% decrease in the onset voltage, which is not a significant drop. As this drop is comparable to the fluctuations as the thickness of PTFE changes and is also not significant enough to require a change in a power source, this can be ignored as a source of error. 

A notable point of uncertainty comes from the way in which equations \ref{eq:onset_voltage_long} and \ref{eq:onset_field} are used in conjunction with the computational simulations. This study assumed that the parameters obtained from these equations were accurate, however, steps were taken to ensure that this is not an unreasonable assumption. As mentioned in the description of the onset voltage model, the electric field strength generated using the results of equation \ref{eq:onset_voltage_long} corresponded to the result calculated by equation \ref{eq:onset_field}, which indicates the values obtained are consistent with each other. And as previously described, when the conditions of a known experiment are replicated, the results match those produced experimentally, indicating that the results are consistent with real-world experimentation. It should be noted that equation \ref{eq:onset_voltage_long} is not the only available equation for calculating the onset voltage, with some work indicating that the equation depicted in equation \ref{eq:onset_voltage_2} is superior \citep{lajhar_electrospray_2018}. Using this equation to recreate the test done to obtain table \ref{tab:test1} gives the results presented in table \ref{tab:test2}. These electric field strength values obtained average out to \SI{3.26}{\volt \per \meter}, which is a 2.16\% decrease from the previously obtained value. This was decided to be too small to be significant. As such it was decided that this does not present a significant enough source of uncertainty to cast any doubt on the validity of the results.

\begin{equation}
    \label{eq:onset_voltage_2}
     V_o=0.7(\frac{ \gamma D cos{\beta}}{\epsilon_o})^{\frac{1}{2}}ln(\frac{8d}{D})
\end{equation}

 \begin{table}[h]
     \centering
          \caption{Capillary test with updated onset voltages}
     \begin{tabular}{c c c}
     \hline
        Distance to extractor (\SI{}{\milli \meter})  & Potential Difference (\SI{}{\volt})& Electric Field Strength (\SI{}{\volt \per \meter}) \\
        \hline \hline
         1.056 & 1776.3 & \SI{3.28e8}{}\\ \hline
         2.112& 1994 & \SI{3.26e8}{}\\ \hline
         3.076 & 2125.1 & \SI{3.25e8}{}\\ \hline
         4.057 & 2219.2 & \SI{3.24e8}{}\\ \hline
     \end{tabular}
     \label{tab:test2}
 \end{table}


These results all indicate that the test is a precise and repeatable way to determine the onset voltage. However, one factor that needs to be considered further is the effect of the propellant used on the results.

\section{Effect of propellant}

It should be reinforced that all previous tests were done using a TEG/ sodium iodide mixture, which is typically used for experimental work carried out in atmosphere. However, ES thrusters are designed for use in vacuum conditions and frequently use ionic liquids for this purpose. As such, these need to be tested as a propellant, both in order to obtain useful data and in order to test the robustness of the previously described methodology.

While one of the major benefits of this type of computational work is the ease with which parameters can be adjusted in order to examine different scenarios, changing the propellant will require some extra steps. This is due to the fact that not only will the material properties of the propellant in the model need to be changed, but both the geometric properties and the onset electric field will also be affected.

 Changing the material properties is the simplest step. For this paper, the ionic liquid chosen was EMI-BF4, which was assigned a relative permittivity of \SI{16.5}{}, and a conductivity of \SI{1.3}{\siemens \per \meter} \citep{stoppa_conductivity_2010}. However, the density, surface tension, viscosity, dielectric constant and conductivity of the propellant all have an effect on the acceptable flow rates of the fluid, according to equation \ref{eq:min_flow} \citep{higuera_qualitative_2017}. This equation depicts the formula for the minimum acceptable flow rate, wherein $Q_{min}$ is the minimum possible flow rate, $\epsilon$ is the relative permittivity, $\rho$ is the density, $K$ is the conductivity and $\mu$ is the viscosity.
 
 \begin{equation}
     Q_{min}=\frac{\epsilon_o \epsilon \gamma}{\rho K}/(\frac{\epsilon_o \epsilon^3 \rho \gamma^2}{\mu^3 K})^{1/3}
     \label{eq:min_flow}
 \end{equation}

The maximum flow rate is typically treated as 100 times the minimum flow rate, which gives the limits of the possible flow rates \citep{reschke_study_2011, el-faramawy_efficiency_2005}. These values will, in turn, affect the geometry of the Taylor Cone, more specifically the radius at the tip of the cone. The relationship between the flow rate and radius is once again obtained from equation \ref{eq:radius_at_tip}.



Referring back to equation \ref{eq:onset_field}, it can be seen that decreasing $r_t$ will increase the electric field necessary for ES emission to occur. Substituting all known figures will show that changing the propellant will result in the required electric field increasing from \SI{3.2e8}{\volt \per \meter} to \SI{6.09e8}{\volt \per \meter}. Similar to the previous propellant, this value can be confirmed by running the capillary model at a range of distances, although this time with the onset voltages adjusted to account for the new propellant (as per equation \ref{eq:onset_voltage_long}). The results of this test averaged out to \SI{6.19e8}{\volt \per \meter}, which is only a 1.6\% increase from the calculated value. 

While the increase in the onset electric field might seem to imply that a higher onset voltage is required, in reality, this is not the case due to the lower radius at the tip resulting in a higher electric field strength being generated. Recreating the test described for PTFE (table \ref{tab:PTFE_results}) results in fluctuations between \SI{810}{\volt} and \SI{830}{\volt}. The mode value is \SI{830}{\volt}, which indicates that the results are similar to those from the test done with the previous propellant. This change is small enough that it would indicate that there is an insignificant difference between using an ionic liquid in vacuum and using a TEG solution in atmosphere when it comes to the onset voltage.

\begin{figure}[h]
    \centering
    \includegraphics[width=0.7\textwidth,trim={0.1cm 0.1cm 0.1cm 0.1cm},clip]{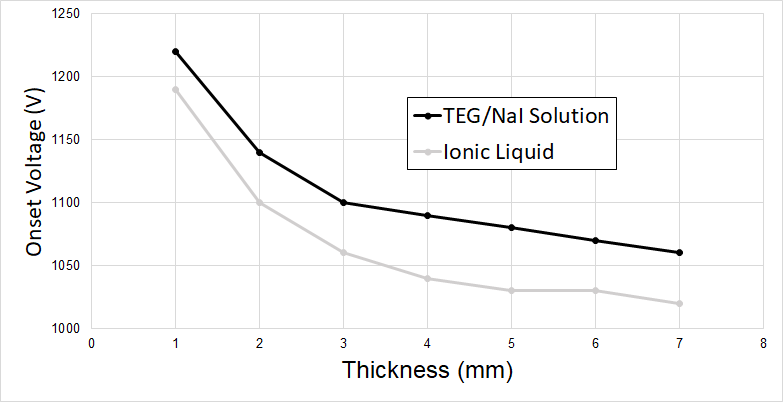}
    \caption{Comparison of onset voltages for PEEK for two different propellants}
    \label{fig:PEEK_onset_voltage}
\end{figure}

Repeating the test described for PEEK (table \ref{tab:PEEK_results}) indicates a more significant change. This is visualised by Fig. \ref{fig:PEEK_onset_voltage}, which compares the onset voltages for the two propellants. Here it can be seen that while both propellants experience a similar trend, the ionic liquid has a moderately lower onset voltage, with this trend slightly increasing as the thickness increases. 


\section{Conclusion}

 These results indicate that under certain conditions, the thickness of the plate can have an effect on the onset voltage, and thus is a factor that needs to be considered when designing flat plate thrusters. In the case of emitters constructed out of PEEK, this change is more pronounced between \SI{1}{\milli \meter} and \SI{3}{\milli \meter} thicknesses, with the onset voltages approaching an equilibrium after these values. As such a \SI{3}{\milli \meter} thick PEEK plate can be considered ideal for manufacturing an emitter of the examined geometry, as it does not have an onset voltage as high as the lower thicknesses, while still minimising the machining time, although of course this decision can vary based on the power supply available. However, the \SI{1}{\milli \meter} thick plate can effectively be considered inefficient, as the notable rise in the onset voltage could have an effect on the power supplies that would be suitable for this mission. 
 
 In the case of the PTFE emitter, the onset voltages were fairly stable across the range of thicknesses, which was consistent with the results observed in \citet{maharaj_propulsion_2021}. It is possible that this is due to its low conductivity value, as the results presented in \citet{maharaj_propulsion_2021} indicate that materials with higher conductivity values experience greater changes in the electric field strength as the thickness changes.
 
 This model provides a comprehensive, robust method for predicting how various parameters will affect the onset voltage. This provides a benefit over traditional analytical methods, which do not easily allow for the geometry to be adjusted. Factors such as the taper, the degree of wetting, and the aspect ratio of the hole can easily be changed, and their effect on the electric field strength and the onset voltage measured. The choice of propellant is the one exception to this rule, as it affects multiple parameters that will need to be calculated and adjusted. However, work done on changing the propellant has shown that it is still possible to simply determine the effect that the propellant has on the onset voltage. 
 
 More specifically, it was found that changing the propellant to one with a lower relative permittivity and higher conductivity resulted in a decrease in onset voltage for most cases. 
 
 
\bibliography{PaperBib}
\end{document}